\documentclass[conference]{IEEEtran}
\IEEEoverridecommandlockouts
\usepackage{cite}
\usepackage{amsmath,amssymb,amsfonts}
\usepackage{algorithmic}
\usepackage{graphicx}
\usepackage{textcomp}
\usepackage{amsmath}
\usepackage{array}
\usepackage{multicol}
\usepackage{rotating}
\usepackage{adjustbox}
\usepackage{multirow}
\usepackage{makecell}
\usepackage{xcolor}
\usepackage{tabularx} % load the package
\usepackage{longtable}
\usepackage{array}
\usepackage{float}
\usepackage{caption}
\usepackage{booktabs} % load the package
\usepackage{multirow} % load the package
\usepackage{siunitx} % load the package
\def\BibTeX{{\rm B\kern-.05em{\sc i\kern-.025em b}\kern-.08em
    T\kern-.1667em\lower.7ex\hbox{E}\kern-.125emX}}
\begin{document}

\title{STM32-Based IoT Framework for Real-Time Environmental Monitoring and Wireless Node Synchronization}

\author{\IEEEauthorblockN{1\textsuperscript{st} Ahmed Faizul Haque Dhrubo}
\IEEEauthorblockA{\textit{Dept.of ECE} \\
\textit{North South University}\\
Dhaka, Bangladesh \\
ahmed.dhrubo@northsouth.edu}
\and
% \IEEEauthorblockN{2\textsuperscript{nd} Sazid Hasan}
% \IEEEauthorblockA{\textit{Dept.of ECE} \\
% \textit{North South University}\\
% Dhaka, Bangladesh \\
% sazid.hasan@northsouth.edu}
% \and
\IEEEauthorblockN{2\textsuperscript{nd} Mohammad Abdul Qayum}
\IEEEauthorblockA{\textit{Dept.of ECE} \\
\textit{North South University}\\
Dhaka, Bangladesh \\
mohammad.qayum@northsouth.edu}
}

\maketitle

\begin{abstract}
The fast pace of technological growth has created a heightened need for intelligent, autonomous monitoring systems in a variety of fields, especially in environmental applications. This project shows the design process and implementation of a proper dual node (master-slave) IoT-based monitoring system using STM32F103C8T6 microcontrollers. The structure of the wireless monitoring system studies the environmental conditions in real-time and can measure parameters like temperature, humidity, soil moisture, raindrop detection and obstacle distance. The relay of information occurs between the primary master node (designated as the Green House) to the slave node (the Red House) employing the HC-05 Bluetooth module for information transmission. Each node displays the sensor data on OLED screens and a visual or auditory alert is triggered based on predetermined thresholds. A comparative analysis of STM32 (ARM Cortex-M3) and Arduino (AVR) is presented to justify the STM32 used in this work for greater processing power, less energy use, and better peripherals. Practical challenges in this project arise from power distribution and Bluetooth configuration limits. Future work will explore the transition of a Wi-Fi communication protocol and develop a mobile monitoring robot to enhance scalability of the system. Finally, this research shows that ARM based embedded systems can provide real-time environmental monitoring systems that are reliable and consume low power.
\end{abstract}


\section{Introduction}
As we know, our world is changing rapidly, and as a result, our everyday way of life is also changing rapidly. The search for the control of fire began an age of human discovery. We now are not only exploring space for knowledge but also translating our technology advances into the ability to occupy and live in space. As our daily way of life has changed, so has the tech that we use every day: televisions, air conditioners, cars, and washing machines are all functional electronic products. Research scientists, engineers, and technologies are continuously working to improve various technologies and minimize the energy consumption of these technologies. In terms of computing, the earliest computers were much larger than today's PCs and laptops, had next to no criteria when solving basic math problems, and were stationary. Nowadays, the size and power of desktops and laptops are almost irrelevant factors, while the technology of laptops was unimaginable years ago and is now a reality. Technology and the internet have seamlessly integrated themselves into everyday tasks and are vital to completing everyday tasks—like burning fire for daily practices. The microcontroller is a critical component necessary for executing the operation of many electronic devices\cite{b1}. 

Microcontrollers, sometimes called embedded controllers, are small microcomputers with a processor, memory, and input/output (I/O) peripherals all on one single chip\cite{b2}. And these pieces work just like a tiny personal computer but are meant to perform specific functions in a larger system. Microcontrollers run without a front end operating system, which is essentially where humans rely on our senses to detect many hazards, while microcontrollers can be embedded into devices that will act as safety devices to signal to the human when a potential hazard is nigh. Microcontrollers are cheap small computing devices that are embedded into devices, and we encounter them every day at work, while at home, most of our appliances have microcontrollers built into them. For example, Arduino and Raspberry Pi are single board computers with simple architectures and good I/O support. Microcontrollers continue to innovate and release many new microcontrollers frequently. The infamous ARM (Advanced RISC Machine) processor based microcontrollers gained rapid traction due to their speed in performance\cite{b3}. They are also low-cost, consume very little power, and produce little heat, making ARM processors perfect for small portable battery powered devices like smartphones, tablets, laptops, and embedded systems.

In ARM architectures, a peripheral device is typically connected to a processor either by explicitly mapping their physical registers into the ARM memory space in addressable memory, using coprocessor space, or through another bus system. Coprocessor access has lower latency; for example, an XScale interrupt controller can be accessed through both memory and coprocessor pathways. ARM processors come in many different architectures or series, such as the Cortex-A series for high-performance applications like gaming and multitasking, and the Cortex-M series for low-power applications where battery life is important\cite{b4}. The architecture and design philosophies of ARM and traditional x86 processors are qualitatively different. x86 processors, found in desktop and laptop computers, are designed with Complex Instruction Set Computing (CISC) architecture, providing the best performance for traditional desktop/laptop General Purpose Computer tasks, whereas ARM processors are based on a Reduced Instruction Set Computing (RISC) architecture, and are designed to be energy efficient and low power for tasks such as mobile devices, embedded systems, and are increasingly being adopted for server and data center workloads\cite{b5}.

\section{Literarture Review}
Valsalan and his colleagues\cite{b6} introduced a portable physiological monitoring system that can constantly observe a patient's heartbeat, body temperature and the environmental conditions of the surrounding room. They stated an operation to continuously and remotely monitoring and control, thereby providing the patients data to a server through Wi-Fi module using a wireless communication solution. In their design, the sensors are placed on the patients body to monitor temperature and heartbeat, while 2 other sensors were placed in the room to measure humidity and ambient temperature. Thus there are 4 sensors in total connected to a control unit moderating the values measured. Pavitha and Ranjith presented an efficient IoT based implementation of effectively monitoring and controlling home appliances through the World Wide Web\cite{b7}. The system enables users to control home appliances utilizing a smart phone and communicates with the use a Wi-Fi and Raspberry Pi Server. The project is low-cost with the use of open-source technologies. The authors were able to incorporate 3 sensors and camera along on a Raspberry Pi computer running an ARM-version of the Debian GNU/Linux OS.

Singh et al.\cite{b8} tested marigolds to find the best-growing conditions for marigolds, using an IoT-based monitoring system to track effects of physical growth parameters (humidity, temperature, soil temperature and moisture, light intensity) on the growth of the plants. The findings showed highest rate of growth at light intensity of 1000 lx–1200 lx (category-2). The authors deployed seven supervised machine learning algorithms. Logistic Regression, Linear SVC, and Gradient Boosting Classifier gave the best accuracy rates of 83.33\%. Suma et al.\cite{b9} proposed an extensive project that included GPS-based remote monitoring, moisture and temperature sensing, intruder detection, security, leaf wetness monitoring, and automated irrigation. The data from many sensors were transmitted over RS232 and processed through a PIC 16F877A microcontroller. The microcontroller has an EEPROM, an electrically erasable programmable read-only memory which allowed it to permanently store pertinent information such and the transmitter codes and the receiving frequencies. An IoT-based environmental changes monitoring framework was introduced by Mosfiqun et al.\cite{b10}, leveraging sensors, microcontrollers, and the Internet of Things technologies. The framework can allow users to monitor indoor and outdoor temperatures, humidity, and harmful gases. The framework included a buck-boost converter as part of the power management, along with a solar cell. The converter was able to manage the variable voltage generated by the solar cell to provide Charing power for the system's battery. The converter took in variable voltages between 3V and 18V and provided regulated 15V to charge the batteries. The device can connect with gateways using Bluetooth, infrared, and Wi-Fi allowing the framework to be utilized across a variety of deployment scenarios. Bhardwaj, Joshi, and Gaur proposed a health monitoring system for rapid diagnosis and treatment\cite{b11}. Their IoT framework aims to alert Physicians of deviations from typical health metrics, such as abnormal heart rate, body temperature, or amount of SpO2. Using commercial health monitoring equipment, Bhardwaj, Joshi, and Gaur determined there were relative errors of 2.89\%, 3.03\%, and 1.05\% in the device's heart rate, body temperature, and SpO2, health metrics, respectively. Their prototype included a microcontroller with a built-in ADC, blood pressure, contactless temperature, and oximeter sensors.

Dhrubo and his team proposed an IoT-based monitoring network for urban and forest area operation\cite{b12}. The design involved deploying a group of mini towers equipped with sensors to collect environmental data and act based on it. Each mini tower will run on a NodeMCU microcontroller with a Wi-Fi module communicating data directly to Firebase. A monitoring tower will contain a Raspberry Pi Pico microcontroller to receive processable data from the mini towers. The developments made will provide a cost-effective, reliable, and autonomous means of responding to threat levels from extreme weather events, ensuring safety for people and vegetation. Rajakumar et al.\cite{b13} proposed the idea of a plant watering system that uses a soil moisture sensor that sends a signal to a water pump to activate when moisture levels drop below a threshold. This system uses ThingSpeak to allow monitoring and control through its compatible iOS showed it has Arduino. An ultrasonic sensor will monitor plant growth, and the status will be sent to the web interface using an IoT module. The watering is performed automatically based on a given time period. Long et al.\cite{b14} presented a holistic monitoring system with FPGA-CPU architecture that monitors conditions and Sub-Synchronous Control Interaction (SSCI) in real-time while logging for post-event analysis. The data logging and network communication are done using a CPU based on the technical limitations of the FPGA. The system utilizes an industrial controller with a 1.91 GHz Quad-Core CPU, 2 GB DRAM, 16 GB storage, Kintex-7 325T FPGA and RJ-45 Gigabit Ethernet port. A UPS is included to maintain data when power loss occurs.

Gupta et al.\cite{b15} designed an air quality monitoring system based on IoT for smart cities. The system measures temperature, humidity, carbon monoxide, LPG, smoke, and particulate matter (PM2.5 and PM10) levels. It used an SDS021 sensor system combined to a Raspberry Pi for measuring the particulate matter value in terms of PPM, rendering the device viable as an all-in-one real-time environmental monitoring device. Sambath et al.\cite{b16} proposed a smart automated plant monitoring system with the use of IoT to provide an irrigation system that offered to lower water consumption. The system was a combination of a ESP8266 Wi-Fi module and an Arduino Uno MCU, and used an LM35 temperature sensor, which do not seem to be present in previous projects. The system allowed for highlighting conditions of the plant and indicates soil purposefulness, the system also analyzed the soil to determine although it may not be appropriate for the plant species in question. Deekshath et al.\cite{b17} used an Arduino UNO and a Wi-Fi module, to process and send the sensed data to the ThingSpeak cloud platform. Their system monitored the parameters every two minutes, using five sensors connected to the Arduino through an amplifier. They used a motor, display and Wi-Fi as components if the overall microcontroller unit (MCU).

\section{Microcontroller Selection Rationale}
In the past, researchers primarily used x86 microcontrollers. However, there is now a growing interest in ARM microcontrollers due to the changing market. The ARM and x86 architectures developed in the 1970s, but they both took very different paths. The x86 architecture was developed by Intel and was adopted very rapidly in IBM personal computers in the early 1980s. This adoption was further accelerated with the extensiveness of Microsoft's DOS operating system, and by the 1990s, x86 had become the dominant architecture for corporate and personal computing. It was nearly a monopolistic standard by the 2000s. Contrastingly, ARM originated as a spin-off of RISC (Reduced Instruction Set Computing) architecture, and was first sold commercially by Acorn Computer Company (UK) with the ARM1 processor in 1985\cite{b18}. ARM's adoption was much slower in the coming decades. The x86 architecture offers very substantial legacy software support, a well-defined and controlled vendor sourcing policy, flexible virtualization, and has become enmeshed within a very large developer ecosystem of hardware and software. x86 is still supported by Windows and every Linux distribution available, which will make the x86 architecture, pragmatically, the best choice for getting the most software compatibility out there.

\begin{table*}[t]
\centering
\caption{Comparative Analysis of Arduino and STM32 Microcontroller Architectures}
\renewcommand{\arraystretch}{1.4} % Improves readability
\begin{tabular}{|p{3cm}|p{5cm}|p{5cm}|}
\hline
\textbf{Aspect} & \textbf{Arduino (ATmega/AVR)} & \textbf{STM32 (ARM Cortex M)} \\ \hline
\textbf{CPU Architecture} & 8-bit AVR (e.g., ATmega328); some use 32-bit SAM3X (e.g., Arduino Due) & 32-bit ARM Cortex-M series (M0–M7), including F0 to H7 families \\ \hline
\textbf{Clock Speed} & 16 MHz (Uno); up to 84 MHz (Due) & From 24 MHz (F0) up to 800 MHz \\ \hline
\textbf{RAM / Flash Memory} & Typically 2 KB RAM / 32 KB Flash (Uno); 96 KB RAM in Due & 64 KB–512+ KB Flash; tens to hundreds of KB RAM depending on model \\ \hline
\textbf{Peripherals and I/O} & GPIO, UART, SPI, I2C; limited ADC/DAC & CAN, USB OTG, camera interface, JPEG decoding, cryptography, timers, DMA \\ \hline
\textbf{Performance} & Suitable for simple tasks; limited in complex or multimedia processing & High performance for real-time, multimedia, or complex embedded tasks \\ \hline
\textbf{Ease of Use / Libraries} & Beginner-friendly IDE and rich library ecosystem & STM32CubeIDE with HAL/LL libraries; more complex setup \\ \hline
\textbf{Coding and Development} & Arduino C/C++; easy syntax, limited low-level access & Requires good C knowledge; supports RTOS and low-level control \\ \hline
\textbf{Community and Support} & Large hobbyist/educational base; many tutorials & Smaller but active community with strong ST documentation \\ \hline
\textbf{Use Cases} & Ideal for education, DIY, and prototyping & Industrial applications, automation, robotics, and embedded systems \\ \hline
\textbf{Hardware Cost and Availability} & Low cost and widely available & Slightly higher cost; Blue Pill/Nucleo boards are affordable options \\ \hline
\textbf{Scalability / Future-Proofing} & Limited by 8-bit architecture & Highly scalable with wide range of performance tiers \\ \hline
\textbf{Code Portability} & Portable but constrained by architecture & STM32duino and HAL support flexible and advanced portability \\ \hline
\end{tabular}
\end{table*}

In contrast, ARM-based systems, owe their energy efficiency to the simple RISC instruction set, and the high levels of integration into ARM chipsets, combined with both ARM processors' physical flexibility, and typically lower price for workloads similar to x86 workloads\cite{b19}. While x86 and ARM will forever be incompatible architectures, advancements have been occurring in cross-platform development, such as Microsoft's new support for both architectures in Windows. In this project an STM32F103C8T6 microcontroller was chosen, a mid-range member of the STM32F103x8 family of ARM Cortex-M3 cores, which is compliant with RISC principles. The STM32F103C8T6 is more commonly found on the Blue Pill development board, which was intended as a low-cost alternative to STMicroelectronics' official STM32 discovery boards.

\begin{figure}[htbp]
  \centering
  \includegraphics[width=1\columnwidth]{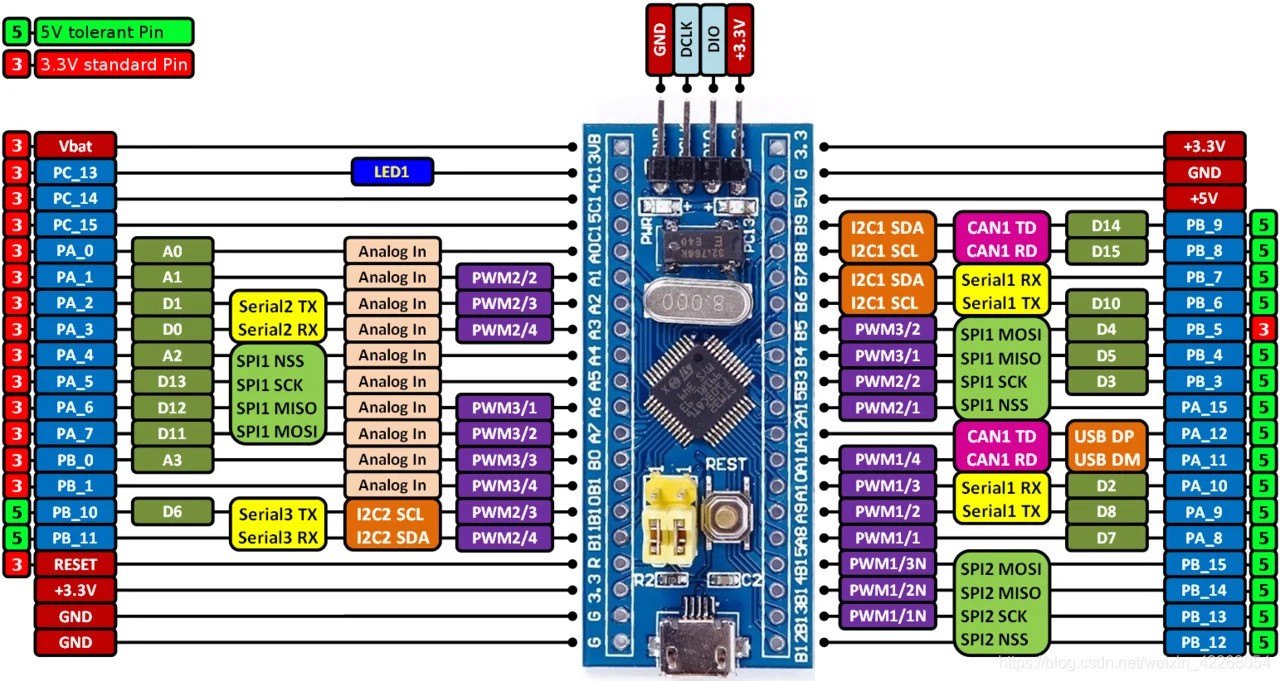} % Adjust the width as needed
  \caption{Pinout diagram of the STM32F103C8T6 Blue Pill microcontroller.}
  \label{fig:j}
\end{figure}

Figure 1 shows PINs and peripheral functions of the STM32F103C8T6 Blue Pill microcontroller. The STM32F103C8T6 can be interfaced using any method they support - it is one of the best boards available for the IoT/edge computing developer, with 37 GPIO pins, 10 analog input channels, and some of the latest communication peripherals, and only physically occupies a small amount of space\cite{b20}. Operated at a high speed of 72 MHz, STM32F103C8T6 supports a very vast range of peripherals e.g. UART, SPI, I2C, Timers, PWM etc. The STM32F103C8T6 has low-power modes of operation, making it a great microcontroller for battery-powered and energy-sensitive applications. Arduino platforms are simple to learn, straightforward, user-friendly, have a well defined developer environment and have a large community support base, making it an ideal tool for education and project learning purposes, but are limited in processing power and scalability. STM32 microcontrollers offer a better option for advanced projects requiring increased processing power, multimedia prowess, and industrial-grade capabilities, at a fraction of the cost. Based on this, it was decided that STM32F103C8T6 is the most suitable option for these research objectives. Table I data is a comparison between STM32 and Arduino. The term processor works is indicated in this table, and all data is collected from the Orient Display article and the Reddit discussion. \cite{b21,b22,b23}.

\section{Methodology}

In this project, we will communicate from one STM32F103C8T6 to another STM32F103C8T6 through the HC-05 bluetooth module. The green house is in a master mood. The Green house is collecting the data and showing the data on his own screen and sending these data simultaneously to the Red house.

\begin{figure*}[t]
\centering
\includegraphics[width=0.8\textwidth]{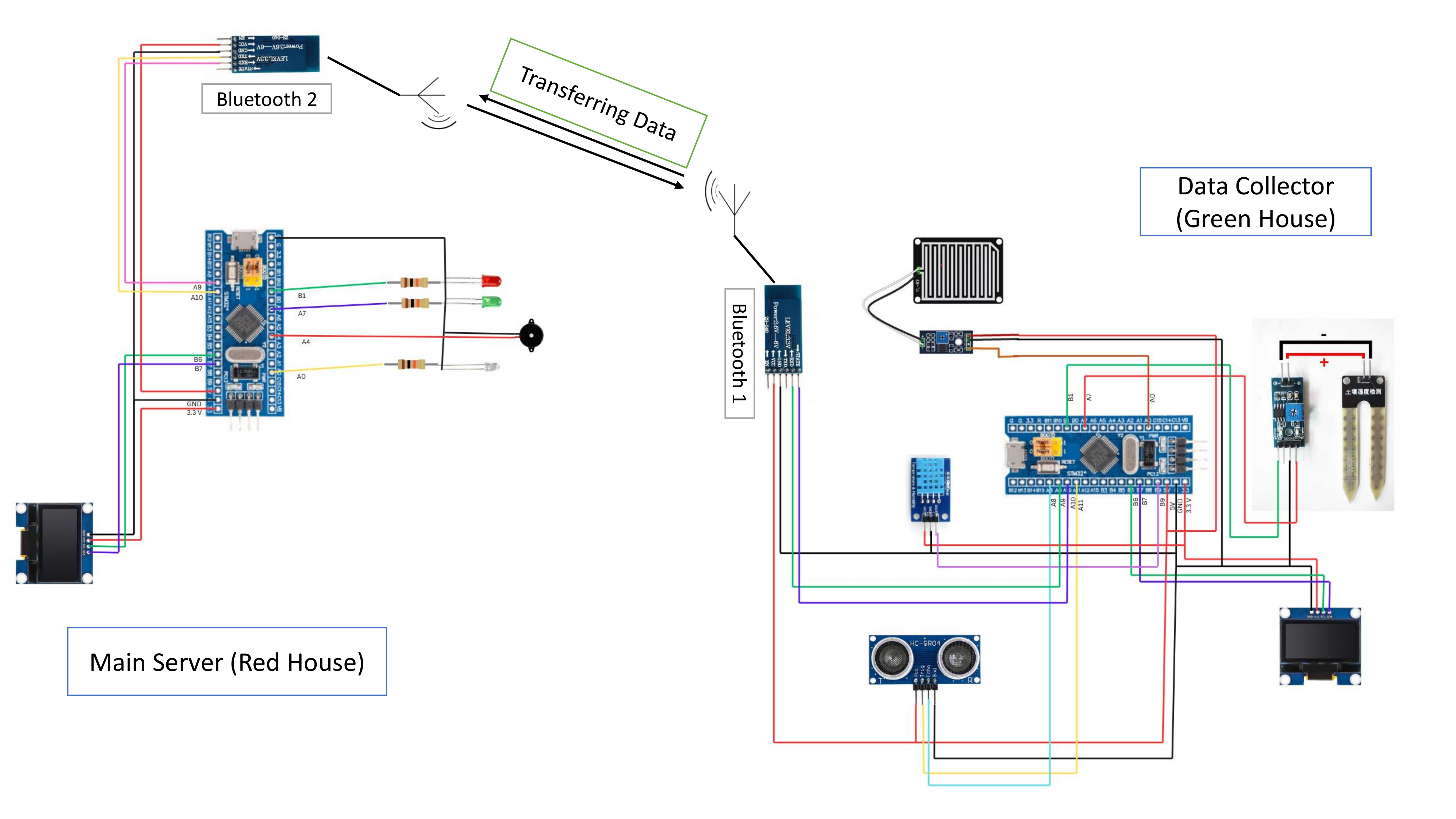}
\caption{Architecture of the Proposed Any Area Alert Monitoring System Using the STM32F103C8T6 Microprocessor.}
\label{fig:jv}
\end{figure*}

Figure 2 presents the backbone design of the whole project. It shows how and which pins the STM32F103C8T6 microprocessor were bridged to the different sensors. In addition, it specifies the pin that is involved in each connection, including the name of the corresponding STM32F103C8T6 microcontroller pin. Within the Red House section, this figure also indicates the resistors involved in blinking the LED lights.

\subsection{Green House}
To implement the transmission system for the Green House, we start with an OLED display and STM32F103C8T6 microcontroller. For connecting the OLED display, we connected the SCL pin to port PB6 of the STM32 and the SDA pin to port PB7. Lastly, the OLED's Vcc is connected to 3.3V of the STM32 and the GND pin to GND of the STM32. In the OLED display, we are displaying temperature, humidity, and distance to an obstacle from the gate of the house. We are using an ultrasonic sensor to measure the distance from an obstacle to the gate in centimeters. The ultrasonic sensor's Vcc and GND pins are connected to STM32's 5V and GND, and the Trig and Echo pins of the ultrasonic sensor are connected to PA11 and PA8 of the STM32. We are using a DHT11 sensor to measure temperature and humidity. The DHT11 sensor's +, – and O pins are connected to 3.3V, GND and PB9 of STM32. 

\begin{figure}[htbp]
  \centering
  \begin{minipage}[b]{0.45\linewidth}
    \centering
    \includegraphics[width=\linewidth]{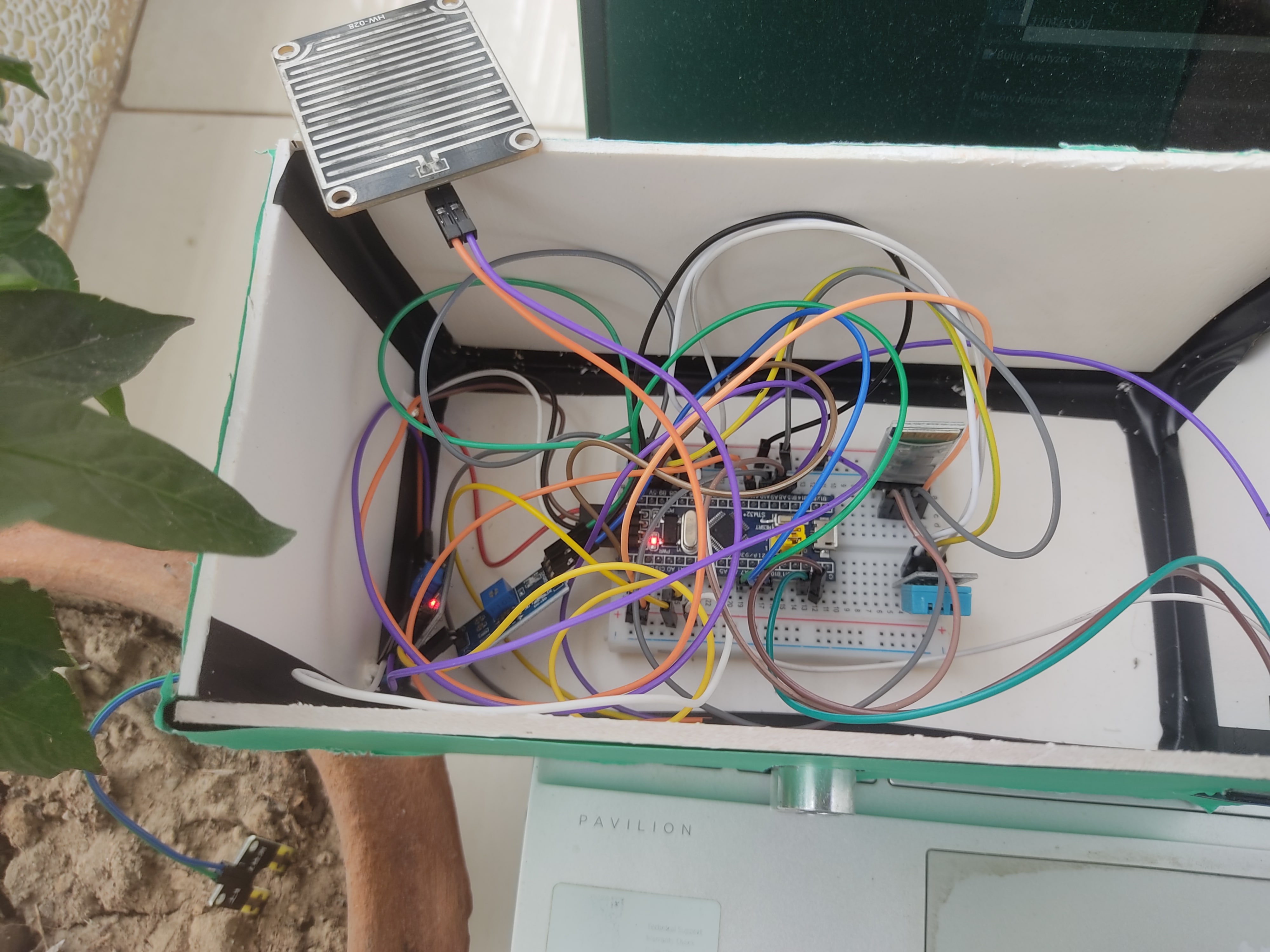}
  
  \end{minipage}
  \hfill
  \begin{minipage}[b]{0.45\linewidth}
    \centering
    \includegraphics[width=\linewidth]{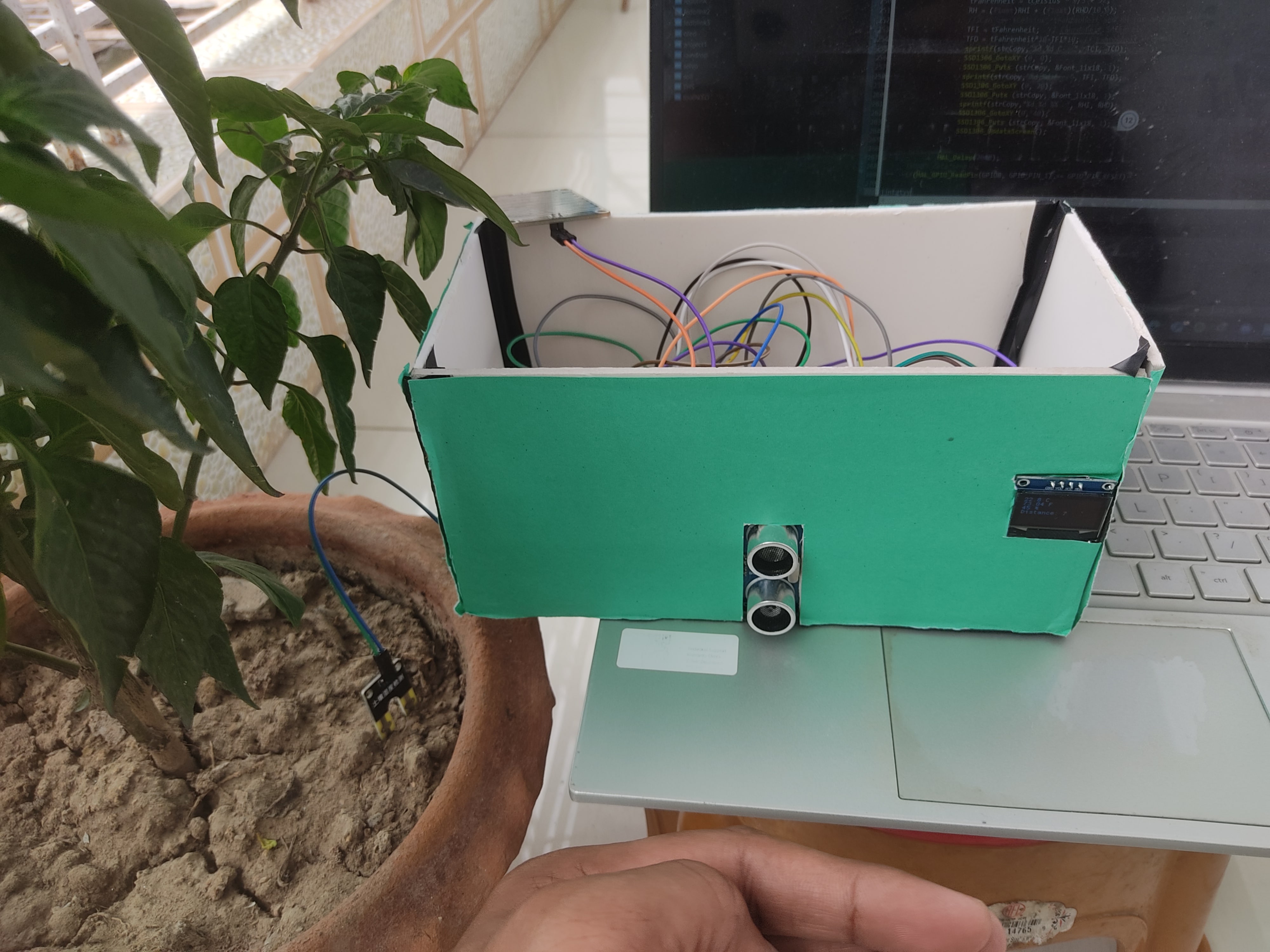}
  
  \end{minipage}
  \caption{Demonstration view of both the interior and exterior of the green house.}
  \label{fig:comparison}
\end{figure}

Figure 3 shows how we designed the interior and exterior of the greenhouse, whose job is to collect data and send it to the red house. Also included is a soil moisture sensor. The Vcc, GND, and A0 pins of this sensor are connected to PA7, GND, and PB1 of the STM32 respectively. The soil moisture sensor indicates the amount of water that is in the soil. A raindrop sensor is also employed which indicates whether water droplets are present or not. Its Vcc, GND, and D0 pins are connected to the 5V, GND, and PA0 ports of the STM32 respectively. The raindrop sensor sends a corresponding signal to STM32 if it detects any water. Lastly, a HC-05 Bluetooth module is on-board. The Vcc and GND of the module are connected to the 5V and GND of STM32 respectively. The Rx and Tx pins of the module are connected to PA9 and PA10 of STM32 respectively. Bluetooth module is configured in operation mode. To properly set the Bluetooth module's operating mode, an Arduino was required because STM32CubeIDE did not provide the ability to directly configure Bluetooth module in operating mode. The master Bluetooth module transmits only one piece of information at a time, and so the clock configuration of the STM32 was set to 72MHz (prescaler value of 71). The USART1 asynchronous mode was enabled with baud rate of 9600.

\subsection{Red House}

The Red House set-up included an STM32F103C8T6 microcontroller interfaced with an OLED display. The SCL pin of the OLED display is wired to port PB6 of the STM32 and the SDA pin is wired to port PB7. The Vcc pin of the OLED display is wired to the 3.3V supply of the STM32 and the GND pin is wired to the GND of the STM32. The OLED display is used to show the temperature and humidity readings as well as the distance of a detected obstacle from the gate, which it receives data from the Green House. Three LEDs are used to show the water level in the soil wired to ports PB1, PA7, and PA0 of the STM32, respectively. A buzzer connected to port PA4 is used to sound an alarm in the event of rainwater detection.

\begin{figure}[htbp]
  \centering
  \begin{minipage}[b]{0.45\linewidth}
    \centering
    \includegraphics[width=\linewidth]{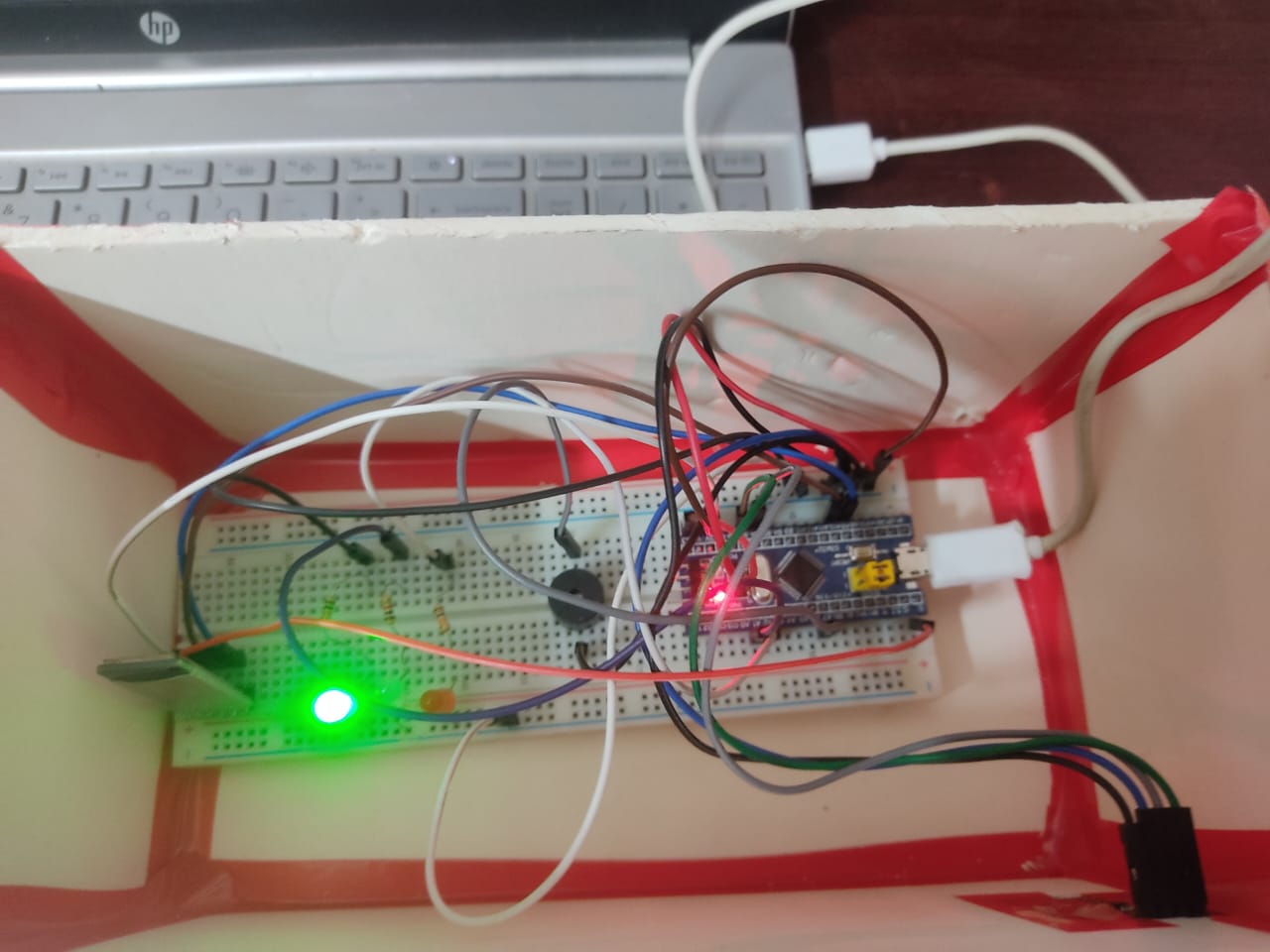}
  
  \end{minipage}
  \hfill
  \begin{minipage}[b]{0.45\linewidth}
    \centering
    \includegraphics[width=\linewidth]{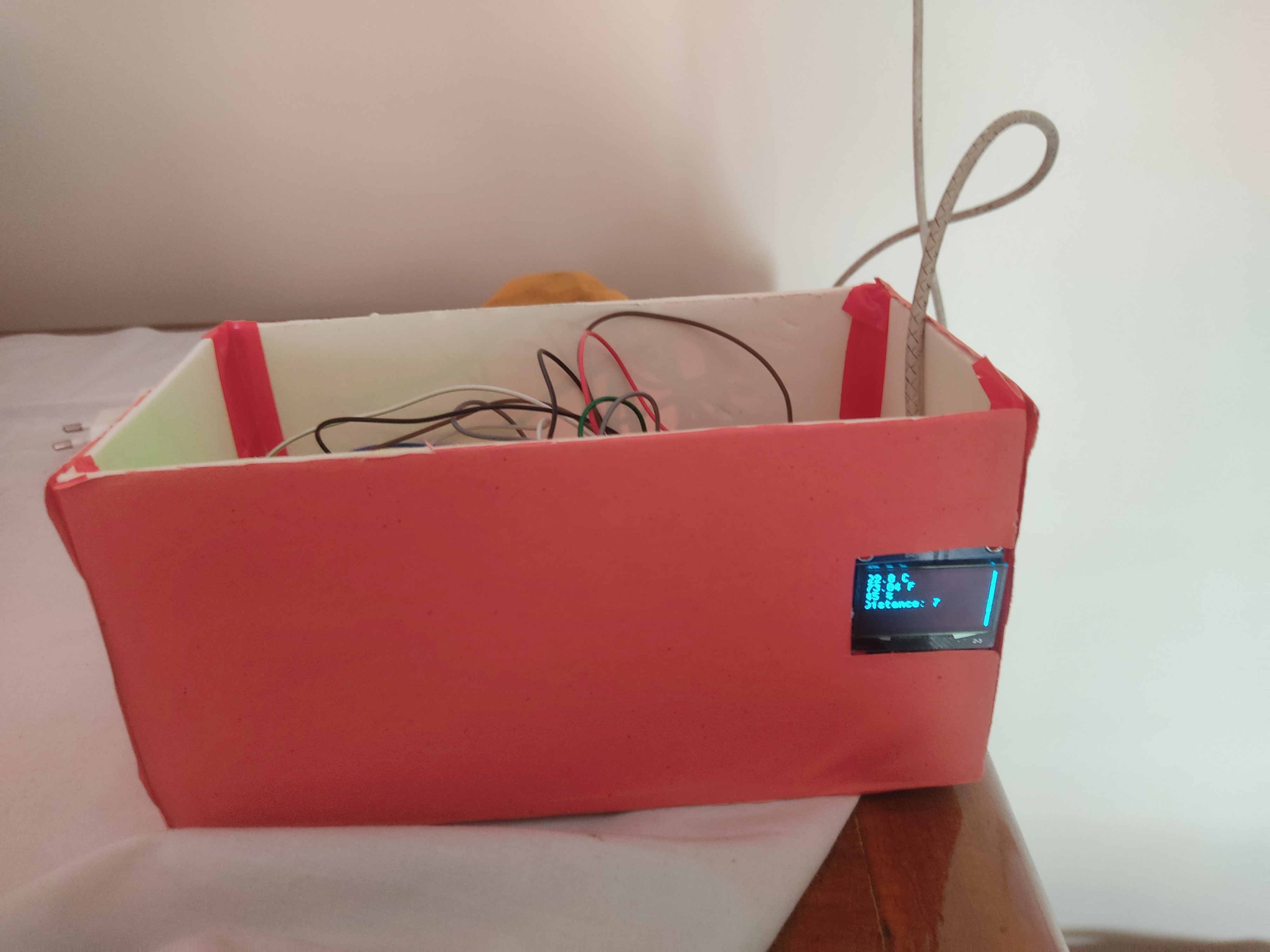}
  
  \end{minipage}
  \caption{Demonstration view of both the interior and exterior of the red house.}
  \label{fig:comparison}
\end{figure}

Figure 4 shows how we designed the interior and exterior of the red house, whose job is to receive data from the greenhouse about what is happening there. There is an HC-05 Bluetooth module integrated into the system. The Vcc and GND pins of the module is connected to the 5 V and GND pins of the STM32 respectively. The Rx and Tx pins of the module are connected to the ports PA9 and PA10 of the STM32. The Bluetooth module is configured to slave mode and can receive one piece of information at a time. Communication will take place in asynchronous mode of USART1 at the baud rate of 9600.

\section{Limitations and Future Work}
There were numerous technical difficulties that were reflected upon and problem solved during the course of this project. The first was in reference to the power supply for the STM32 development platform. At first, several components were all branched to the same 3.3V and GND ports alone. This proved insufficient because the STM32 platform did not have enough power for the OLED display to work. Finally, it was determined that the OLED display could be powered from a different 3.3V and GND port which worked. The second major issue was getting the Bluetooth module to work wirelessly for communication. The module had to be set as a master device, there was no way to configure this master mode option in the STM32CubeIDE. Therefore, an Arduino board was used to set the Bluetooth module as the master device. The Bluetooth module was successful in changing to the correct mode; however, other issues became apparent after its configuration its performance characteristics meaning it can only transfer one data 'pack' at a time. It was also limited by speed for transmission and distance, which were both characteristics affecting real-time monitoring.

Considering these limitations, future improvements will include replacing the Bluetooth with a stronger wireless communication protocol such as Wi-Fi. There are benefits for using Wi-Fi including faster transfer rates, a larger operational range, and better compatibility with web services. Additionally, the ultimate aim is to reconfigure the stationary monitoring system into a mobile unit that is potentially either a standard wheeled robot, or similar. This mobile monitoring robot (MMR) could navigate through the greenhouse, collect data from specified locations, and wirelessly send data to the cloud.

\section{Conclusion}
Due to the rapid growth of science and increase adoption of modern technology, intelligent systems have found increased use in many applications, with environmental monitoring seeing real usage lately in intelligent systems, which has also influenced the supporting hardware platforms. Historically, x86-based microcontrollers were used due to their availability and ease of use, but for this project we are adopting an ARM-based STM32 microcontroller. The architecture of the STM32 is more challenging than x86; however, the advantages of using an ARM-based microcontroller are significant in performance and energy use.
 
By working with the STM32, it became clear to us that ARM-based microcontrollers are better suited for real-time applications which require a higher processing capacity and energy-efficient options. Although the learning curve at the beginning is steep, the professional development tools that are available, are very useful in reducing any difficulties. STM32CubeIDE, STM32CubeMX, and Keil all provide professional-rank development tools designed to facilitate development inclusive of project generation, hardware abstraction layer (HAL), and in-circuit debugging, all without added costs. Since this project represents an initial inquiry into embedded environmental monitoring, we were only able to use a few sensors in this initial stage of the process. In future iterations we will not only deploy more sensors, we will also design the system to support new opportunities and broaden their use, therefore increasing their value and usefulness in applications.

\vspace{12pt}
\color{red}

\end{document}